\documentclass[conference]{IEEEtran}
\IEEEoverridecommandlockouts
\usepackage{cite}
\usepackage{amsmath,amssymb,amsfonts}
\usepackage{algorithmic}
\usepackage{graphicx}
\usepackage{textcomp}
\usepackage{xcolor}
\usepackage{booktabs}
\usepackage{array}
\usepackage[utf8]{inputenc}
\usepackage[T1]{fontenc}
\usepackage[english]{babel}
\usepackage[hidelinks]{hyperref}

\def\BibTeX{{\rm B\kern-.05em{\sc i\kern-.025em b}\kern-.08em T\kern-.1667em\lower.7ex\hbox{E}\kern-.125emX}}

\begin{document}

\title{Multi-Label 12-Lead ECG Classification on the PTB-XL Dataset: A Comparative Evaluation of Deep Learning Architectures and Heterogeneous Ensemble Approaches}

\author{
\IEEEauthorblockN{Yunus Emre Mert}
\IEEEauthorblockA{\textit{Department of Mechatronics Engineering} \\
\textit{Yildiz Technical University}\\
Istanbul, Turkey \\
ynsemremert@gmail.com}

\and
\IEEEauthorblockN{Ece Akdoğan}
\IEEEauthorblockA{\textit{Robert College}\\
Istanbul, Turkey \\
eceakdogan2608@gmail.com}

\and
\IEEEauthorblockN{Hüseyin Üvet}
\IEEEauthorblockA{\textit{Department of Mechatronics Engineering} \\
\textit{Yildiz Technical University}\\
Istanbul, Turkey \\
huvet@yildiz.edu.tr}
}
\maketitle

\begin{abstract}
This study aimed to compare the performance of different deep learning architectures and heterogeneous ensemble learning approaches for multi-label 12-lead ECG classification on the PTB-XL dataset. Five different models, namely 1D-ResNet18, Bidirectional Mamba, xLSTM, CWT-ViT-KAN, and the pre-trained ECGFounder, were evaluated. Utilizing the recommended split structure of the PTB-XL dataset, folds 1–8 were allocated as the training set, fold 9 as the validation set, and fold 10 as the independent test set. In addition to the individual models, three different ensemble approaches were investigated: Equal-Weight Soft Voting, Validation-Weighted Soft Voting, and stacking. Among the individual models, the highest performance was achieved by ECGFounder, with a Macro AUROC of 0.930 and a Macro AUPRC of 0.823. For the ensemble models, the highest values in the primary macro performance metrics were obtained by the stacking approach, achieving a Macro AUROC of 0.936, a Macro AUPRC of 0.836, and a Macro F1 of 0.763. The highest subset accuracy of 0.630 was achieved using the Validation-Weighted Soft Voting method. The findings indicate that heterogeneous ensemble models, which combine different representation learning approaches, can provide additional performance improvements over individual models in multi-label ECG classification.
\end{abstract}


\section{INTRODUCTİON}

Cardiovascular diseases (CVD) are currently among the most significant health problems globally. They rank among the leading causes of death worldwide \cite{wagner2022}. Therefore, early diagnosis studies are becoming increasingly important. The primary diagnostic method is the electrocardiogram (ECG), a non-invasive technique \cite{strodthoff2021}. However, the accurate interpretation of ECGs requires clinical experience and expertise. Furthermore, it is susceptible to human error, and the interpretation of results may exhibit variability \cite{garg2024}. This has increased the utilization of machine learning and deep learning approaches in medical studies. The deep learning models currently used demonstrate cardiologist-level performance in ECG interpretation \cite{strodthoff2021}.

Research on ECG classification has evolved rapidly, progressing from machine learning to deep learning architectures. Algorithms such as decision trees, logistic regression, and random forests were the most commonly used algorithms in machine learning \cite{garg2024}. However, these models were insufficient in capturing the complex and non-linear patterns present in ECG data \cite{garg2024}. In contrast, deep learning models have been preferred over machine learning methods due to their automated learning capabilities \cite{garg2024}. In particular, Convolutional Neural Networks (CNNs), including ResNet- and Inception-based architectures, are prominent architectures that have demonstrated high performance on ECG signals \cite{strodthoff2021}.

One of the most significant developments supporting ECG research is PTB-XL, one of the largest open-source clinical 12-lead ECG datasets published to date \cite{wagner2022}. The scale and pathological diversity of this dataset enable the training of deep learning algorithms and facilitate the study of potential real-world patient scenarios \cite{wagner2022}.

It is known that successful results have been achieved in ECG classification using standard CNN- and LSTM-based models. At the same time, areas such as the modeling of long sequences, the capture of fine-grained features in the time-frequency domain, and the utilization of large-scale pre-trained models remain open to further development.

For this reason, the objective of this study is to evaluate the performance of 1D-ResNet18, Bidirectional Mamba, xLSTM, Continuous Wavelet Transform - Vision Transformer - Kolmogorov-Arnold Network (CWT-ViT-KAN), and the pre-trained ECGFounder models on the PTB-XL dataset, and to investigate the potential contribution of heterogeneous ensemble learning.

\section{MATERIALS AND METHODS}

\subsection{Dataset and Preprocessing}

\begin{figure}[htbp]
    \centering
    \includegraphics[width=0.8\linewidth]{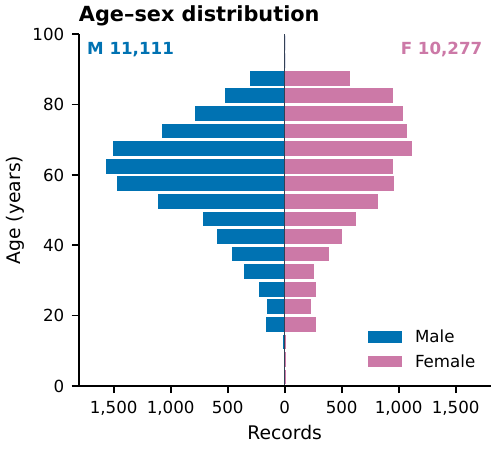}
    \caption{Demographic pyramid illustrating the age and gender distribution of patients in the PTB-XL dataset.}
    \label{fig:age_sex_dist}
\end{figure}

\begin{figure}[htbp]
    \centering
    \includegraphics[width=0.8\linewidth]{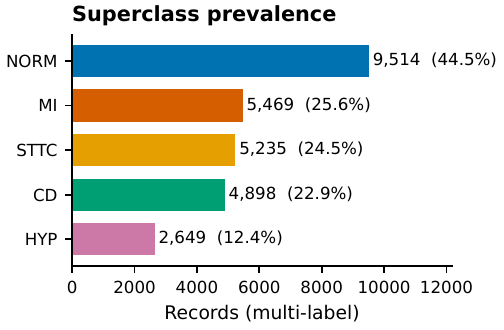}
    \caption{Distribution of the five primary diagnostic superclasses within the PTB-XL dataset. As ECG records may contain multiple labels, the sum of the class frequencies may exceed the total number of unique records.}
    \label{fig:superclass_dist}
\end{figure}

In this study, the PTB-XL v1.0.3 dataset, which is frequently used in contemporary ECG research, was utilized. The dataset comprises 21,799 clinical 12-lead ECG records obtained from 18,869 patients, each with a duration of 10 seconds at sampling frequencies of both 100 Hz and 500 Hz. After excluding records that did not correspond to any of the five diagnostic superclasses considered in this study, 21,388 ECG records remained for the final analysis. Of these records, 11,111 (52.0\%) were obtained from male patients and 10,277 (48.0\%) from female patients (Figure \ref{fig:age_sex_dist}). There are five main diagnostic superclasses present in this data: Normal ECG (NORM), Myocardial Infarction (MI), ST/T Change (STTC), Conduction Disturbance (CD), and Hypertrophy (HYP). The distribution of the classes is illustrated in Figure \ref{fig:superclass_dist} \cite{wagner2022}.

For the 1D-ResNet18, Bidirectional Mamba, xLSTM, and CWT-ViT-KAN models, the 100 Hz records of the PTB-XL dataset were utilized. The inputs for these models were structured as matrices with dimensions of $12 \times 1000$, representing the 10-second 12-lead ECG records. Conversely, to align with the native structure of the pre-trained architecture, the ECGFounder model was subjected to fine-tuning using the native 500 Hz records of the PTB-XL dataset. Consequently, the inputs for ECGFounder were formed from 10-second, 12-lead ECG signals with dimensions of $12 \times 5000$. The same training, validation, and test splits were used for all models.

\subsection{Label Generation}

The labels in the PTB-XL 1.0.3 dataset were defined using SCP-ECG statements generated by cardiologists. The \texttt{scp\_codes} present in each record were categorized into five main diagnostic superclasses, namely NORM, MI, STTC, CD, and HYP, by considering only the statements designated as \texttt{diagnostic = 1} in the \texttt{scp\_statements.csv} file. Form- and rhythm-related codes were not included in the evaluation, and no additional threshold was applied for likelihood. Recurring diagnostic classes within the same record were deduplicated, and records that did not correspond to any diagnostic superclass were excluded from the study. Finally, the remaining labels were encoded as five-dimensional multi-label (\textit{multi-hot}) vectors according to a fixed class order \cite{wagner2022}.

\subsection{Deep Learning Architectures Utilized}

In this study, five distinct deep learning architectures, encompassing convolutional, state-space, recurrent, and transformer-based approaches, were employed and adapted to one-dimensional (1D) ECG waveforms to capture the temporal and morphological patterns of cardiac signals.

\subsubsection{1D-ResNet18 (Baseline Model)}

This architecture is an artificial intelligence model based on the residual learning principle, which was developed to mitigate the performance degradation that can occur as the number of layers increases in deep neural networks. Shortcut connections enable each block in the network to learn a residual correction based on its input, while facilitating a more efficient propagation of gradients throughout the network. In this study, the standard ResNet18 architecture was adapted to temporal ECG signals by utilizing 1D convolutions. This architecture served as a robust baseline model for capturing local morphological features, such as QRS complexes and wave transitions, while maintaining low computational costs \cite{he2016}.

\subsubsection{Bidirectional Mamba (Selective State-Space Models)}

The Mamba architecture, developed on the basis of State-Space Models to efficiently model long sequences, offers linear time complexity while avoiding the quadratic computational cost associated with Transformer architectures. The most significant innovation of Mamba is its "selective" mechanism, which defines state transition parameters as functions of the input data rather than keeping them constant. The Bidirectional Mamba (Vim) variant utilized in this study scans the signal in both forward and backward directions and integrates the information obtained from the two passes. In datasets such as ECG where the entire signal is pre-existing, the concurrent utilization of context preceding and succeeding any given point can contribute to a more comprehensive modeling of diagnostic information \cite{gu2023, zhu2024}.

\subsubsection{xLSTM (sLSTM / mLSTM)}

xLSTM is an extended architecture proposed to overcome the limitations of conventional Long Short-Term Memory (LSTM) networks, such as the inability to revise past decisions, reliance on a single cell state, and parallelization difficulties. In xLSTM, the conventional design is updated with exponential gating and two novel memory cells: sLSTM, which features a scalar memory offering enhanced mixing capacity, and mLSTM, which possesses a fully parallelizable matrix memory. The sequential utilization of these cells in blocks is intended to concurrently process short-term morphological details and longer-term rhythm patterns within ECG records \cite{beck2024}.

\subsubsection{CWT-ViT-KAN}
In this architecture, classical signal processing techniques in the time-frequency domain are combined with Kolmogorov-Arnold Network (KAN) and Vision Transformer (ViT) structures. Initially, 1D ECG leads are decomposed into time-frequency components and converted into 2D scalograms by the Continuous Wavelet Transform (CWT). Subsequently, these scalograms are processed utilizing the Wav-KAN approach, which incorporates learnable wavelet functions on the edges of the network rather than employing fixed activation functions at the nodes. In the final stage, the data is divided into patches by the ViT, and global relationships are modeled through a self-attention mechanism without the need for convolutional layers. This integrated architecture is designed to jointly represent fine- and coarse-scale features in the time-frequency domain, in addition to long-range dependencies \cite{liu2024}, \cite{bozorgasl2024}, \cite{dosovitskiy2020}.

\subsubsection{ECGFounder (Fine-Tuning)}

ECGFounder is a large-scale, convolutional foundation model developed for direct ECG interpretation. The model was pre-trained on over 10 million expert-labeled records obtained from the Harvard-Emory ECG Database, encompassing 150 distinct diagnostic labels. It is designed to support both single-lead wearable device data and standard 12-lead records. In this study, ECGFounder was not trained from scratch; instead, it was implemented utilizing a transfer learning approach. The pre-trained weights of the model were subjected to fine-tuning to align with the features specific to the target classification problem within the PTB-XL dataset. Fine-tuning was executed using the native 500 Hz records of the PTB-XL dataset \cite{li2024}.

\subsection{Heterogeneous Ensemble Learning Strategy}

Ensemble learning is a machine learning approach in which the predictions of multiple base models are combined to enhance generalization performance. The concurrent utilization of models with distinct inductive biases and representation learning mechanisms can enable the capture of features that might not be fully captured by a single architecture \cite{zhou_ensemble}.

In this study, heterogeneous ensemble models were constructed using the probability outputs of five individual architectures: 1D-ResNet18, Bidirectional Mamba, xLSTM, CWT-ViT-KAN, and ECGFounder. These models represent convolutional, state-space, recurrent, time-frequency/transformer-based, and ECG-specific pre-trained foundation model approaches, respectively. Regardless of their validation performances, all five individual models were retained as Base Learners in the ensemble analysis, and no minimum \textit{Macro F1} criterion was applied to exclude any model.

Three ensemble strategies were evaluated. First, an Equal-Weight Soft Voting approach was applied by averaging the class-based probability predictions of the five models with equal weights. Second, a Validation-Weighted Soft Voting approach was implemented, wherein the model weights were determined solely based on performance on the validation set and fixed prior to the final evaluation. Third, a Stacking Ensemble was constructed by providing the class-based probability outputs of the five base models as input features to a logistic regression Meta-Classifier. The Meta-Classifier was trained utilizing only the predictions from the validation set and their corresponding ground truth labels, and was subsequently applied to the independent base model predictions. Ensemble configuration, model weighting, and Meta-Classifier training were executed in strict isolation to prevent information leakage; the test set was reserved exclusively for the final evaluation.

In all ensemble methods, class-specific decision thresholds were determined using the validation set and fixed prior to evaluation on the separate test set.

\subsection{Experimental Setup and Evaluation Criteria}

To ensure a consistent comparison among the deep learning models, identical data subsets, label definitions, model selection criteria, and evaluation procedures were utilized across all experiments.The labels NORM, MI, STTC, CD, and HYP were employed for classification. To align with standard literature and ensure a rigorous and reproducible comparison, the recommended 10-fold stratified split of the PTB-XL dataset was applied as described. Folds 1–8 were designated as the training set, fold 9 as the validation set, and fold 10 as the test set. This splitting procedure prevented the distribution of records belonging to the same patient across different data partitions, thereby eliminating the risk of data leakage.

Instead of employing a fixed threshold across all classes, class-specific decision thresholds were determined exclusively on the validation set. For each diagnostic superclass, the threshold value maximizing the validation F1 score was selected and fixed prior to evaluation on the independent test set.

\subsubsection{Training Protocol and Optimization}

To account for class imbalance, BCEWithLogitsLoss with class-specific positive weights (\textit{pos\_weight}) was utilized. The positive weights were calculated exclusively from the class distribution within the training set, as the ratio of the number of negative samples to the number of positive samples for each class. Validation and test labels were not utilized in the calculation of these weights.

AdamW was utilized as the optimizer for all models. The batch size and learning rate were permitted to vary according to the computational requirements and optimization characteristics of each architecture. A maximum of 50 training epochs was applied, and early stopping was implemented based on the validation \textit{Macro AUPRC} value. Training was terminated if no improvement was observed for 7 consecutive epochs, and the checkpoint with the highest validation \textit{Macro AUPRC} was saved for subsequent evaluations.

Data augmentation and mixup were not applied in these experiments. In accordance with the input format expected by the pre-trained Net1D architecture, ECGFounder was fine-tuned utilizing the native 500 Hz records (12 leads, 10 seconds; $12 \times 5000$) of the PTB-XL dataset. As part of the preprocessing, a 0.5–40 Hz band-pass filter was applied to the signals. Subsequently, for lead-wise normalization, statistics calculated exclusively from the training set were utilized, and the identical parameters were applied unaltered to the validation and test sets. Finally, an intra-record global z-score standardization was performed across all leads and time samples for each record. No resampling or linear interpolation was applied.

\subsubsection{Model Hyperparameters}

The hyperparameters and structural characteristics specific to the base model are as follows:

\begin{itemize}

\item \textbf{1D-ResNet18:} A batch size of 64, an initial learning rate (LR) of $1 \times 10^{-3}$, and a weight decay (WD) of $1 \times 10^{-4}$ were utilized.

\item \textbf{Bidirectional Mamba:} A batch size of 32, a learning rate (LR) of $5 \times 10^{-4}$, and a weight decay (WD) of $1 \times 10^{-4}$ were utilized. The model dimension (\texttt{d\_model}) was set to 128, and the state-space dimension (\texttt{d\_state}) was set to 16.

\item \textbf{xLSTM:} A batch size of 32, an LR of $5 \times 10^{-4}$, and a WD of $1 \times 10^{-4}$ were specified. The architecture incorporates sequential sLSTM and mLSTM components within the xLSTM framework to model both local temporal dependencies and longer-range sequence information.

\item \textbf{CWT-ViT-KAN:} A batch size of 16, an LR of $3 \times 10^{-4}$, and a WD of $1 \times 10^{-4}$ were used. Time-frequency representations were obtained by applying a Morlet wavelet-based Continuous Wavelet Transform with 32 scales to the ECG signals. The resulting scalograms were processed according to the model-specific preprocessing pipeline, resized to an input resolution of $224 \times 224$, and fed into the ViT-tiny backbone with a patch size of 16.

\item \textbf{ECGFounder (Fine-Tuning):} A batch size of 16, a learning rate (LR) of $1 \times 10^{-4}$, and a weight decay (WD) of $1 \times 10^{-5}$ were utilized. The model was initialized with pre-trained weights on ECG data and subjected to fine-tuning on the native 500 Hz records of the PTB-XL training set.

\end{itemize}

To ensure reproducibility, a fixed random seed value (seed = 42) was used in all experiments. The individual models were evaluated using the identical data splits, label definitions, and evaluation protocol. The heterogeneous ensemble strategies were subsequently constructed utilizing the probability outputs obtained from these models.

\subsubsection{Evaluation Criteria}

Model performance was evaluated using both threshold-independent and threshold-dependent metrics. Due to the class imbalance within the dataset, \textit{Macro AUPRC} was designated as the primary model selection criterion and was utilized as the primary performance metric alongside \textit{Macro AUROC}.

\textit{Macro F1}, \textit{Micro F1}, \textit{Macro Precision}, \textit{Macro Recall}, and \textit{Macro Specificity} were selected as secondary metrics. For a detailed analysis of the models, these metrics were also calculated separately on a per-class basis for the NORM, MI, STTC, CD, and HYP classes.

Given that the study entails a multi-label classification task, the \textit{Subset Accuracy} metric, which indicates the rate at which all five labels are simultaneously predicted correctly for each ECG record, and the \textit{Hamming Loss} metric, which represents the proportion of incorrect classification decisions, were also incorporated into the analysis.

\section{RESULTS}

\begin{table*}[t]
\centering
\caption{Performance comparison of individual deep learning models and heterogeneous ensemble strategies on the PTB-XL independent test set (fold 10).}
\label{tab:performance_comparison}
\resizebox{\textwidth}{!}{%
\begin{tabular}{lccccccccc}
\toprule
\textbf{Model} &
\textbf{Macro AUROC} &
\textbf{Macro AUPRC} &
\textbf{Macro F1} &
\textbf{Micro F1} &
\textbf{Macro Precision} &
\textbf{Macro Recall} &
\textbf{Macro Specificity} &
\textbf{Subset Accuracy} &
\textbf{Hamming Loss} \\
\midrule
1D-ResNet18 &
0.922 & 0.797 & 0.738 & 0.766 & 0.701 & 0.785 & 0.891 &
0.573 & 0.128 \\
Bidirectional Mamba &
0.923 & 0.816 & 0.739 & 0.768 & 0.703 & 0.780 & 0.894 &
0.583 & 0.126 \\
xLSTM &
0.910 & 0.782 & 0.718 & 0.757 & 0.706 & 0.736 & 0.894 &
0.574 & 0.130 \\
CWT-ViT-KAN &
0.874 & 0.709 & 0.664 & 0.704 & 0.629 & 0.706 & 0.856 &
0.497 & 0.165 \\
ECGFounder (Fine-Tuning) &
0.930 & 0.823 & 0.749 & 0.787 & 0.723 & 0.778 & 0.898 &
0.602 & 0.116 \\
\midrule
\textbf{Equal-Weight Soft Voting} &
0.933 & 0.829 & 0.759 & 0.789 & 0.741 & 0.779 & 0.911 &
0.625 & 0.112 \\
\textbf{Validation-Weighted Soft Voting} &
0.933 & 0.830 & 0.760 & 0.790 & 0.744 & 0.778 & 0.912 &
\textbf{0.630} & 0.111 \\
\textbf{Stacking Ensemble} &
\textbf{0.936} & \textbf{0.836} & \textbf{0.763} &
\textbf{0.795} & \textbf{0.747} & 0.782 & 0.909 &
0.625 & \textbf{0.110} \\
\bottomrule
\end{tabular}%
}
\end{table*}

\subsection{Comparison of Model Performances}

An analysis of the results reveals that ECGFounder demonstrates the strongest performance, achieving the highest values for Macro AUROC (0.930) and Macro AUPRC (0.823). The ECGFounder model is followed by Bidirectional Mamba and 1D-ResNet18. CWT-ViT-KAN exhibits the lowest performance among them. The additional computational and representational costs of the more complex pipeline, which integrates a CWT-based two-dimensional transformation with ViT and KAN components, did not translate into a performance improvement under these experimental conditions, resulting in a less favorable complexity-performance trade-off for the model.

Heterogeneous ensembles demonstrated more consistent performances compared to the individual models. The Stacking Ensemble achieved the best results in the primary discriminative metrics on the test set, attaining a Macro AUROC of 0.936, a Macro AUPRC of 0.836, and a Macro F1 score of 0.763. Notably, the increase in the Macro AUPRC value from 0.823 in the best-performing individual model to 0.836 demonstrates that integrating the complementary predictions of diverse architectures contributes to robust performance against class imbalance.

\subsection{Effect of Ensemble Learning on Subset Accuracy}

On the independent test set, ECGFounder achieved the highest Subset Accuracy (0.602) and the lowest Hamming Loss (0.116) among the individual models. The ensemble models provided further improvements in exact label matching; the highest Subset Accuracy of 0.630 was attained using the Validation-Weighted Soft Voting method. In contrast, the lowest Hamming Loss of 0.110 was observed in the Stacking Ensemble. Thus, while the Stacking Ensemble excelled in the primary macro-level performance metrics, Validation-Weighted Soft Voting yielded the best result in the stricter exact match metric, which requires all labels to be correctly predicted simultaneously.

\section{DISCUSSION}

The evaluations conducted on the PTB-XL test set have clarified the effects of various representation learning mechanisms and heterogeneous ensemble strategies on multi-label ECG classification. While ECGFounder stood out among individual models—with Bi-Mamba and 1D-ResNet18 remaining competitive—heterogeneous ensemble approaches yielded performance improvements, particularly in macro-level metrics.

\subsection{Model Performances and Architectural Implications}

The highest values achieved by ECGFounder, which leverages large-scale pre-training, in the Macro AUROC (0.930) and Macro AUPRC (0.823) metrics demonstrate the power of transfer learning and its model-specific high-resolution input structure in ECG classification. Conversely, the fact that the relatively simpler, task-specific 1D-ResNet18 architecture yielded the best result in macro sensitivity indicates that foundational architectures still maintain their relevance for certain metrics.

It is noteworthy that Bidirectional Mamba (Bi-Mamba), trained from scratch, performed at a level closely approaching that of ECGFounder (0.923 Macro AUROC, 0.816 Macro AUPRC). Selective state-space models achieved competitive performance in capturing long-range dependencies within 10-second ECG sequences without requiring pre-training.

Increasing model complexity does not consistently result in enhanced performance. CWT-ViT-KAN, which transforms ECG signals into a 2D time-frequency space utilizing CWT and processes them with ViT and KAN, yielded the lowest results across all metrics (Macro AUROC: 0.874, Macro AUPRC: 0.709, Macro F1: 0.664). This outcome indicates that under the present experimental conditions, more complex time-frequency representations did not translate into a performance improvement.

\subsection{Subset Accuracy Analysis of the Ensemble Model}

The performance improvements observed during validation were also reflected on the independent test set. The Stacking Ensemble achieved the best overall performance across the primary macro metrics, with a Macro AUROC of 0.936, a Macro AUPRC of 0.836, and a Macro F1 score of 0.763. The meta-classifier was trained only on validation predictions, while the test set remained untouched until final evaluation. This design reduces the risk of test-set leakage and suggests that the improvement may result from combining complementary predictions from the base models.

A distinct dynamic was observed regarding Subset Accuracy, which measures exact multi-label matching. While the Stacking Ensemble approach attained a value of 0.625, the simpler Validation-Weighted Soft Voting method achieved the highest score of 0.630. These results indicate that heterogeneous ensembles, which leverage the complementary characteristics of diverse representation learning approaches, can offer potential advantages.

\subsection{Limitations}

In this study, the ECGFounder model was fine-tuned utilizing the native 500 Hz records of the PTB-XL dataset, whereas the other individual models were trained on 100 Hz records. Therefore, the performance differences among the models should not be attributed solely to architectural characteristics or the effect of pre-training. Furthermore, the validation set was utilized across multiple stages, such as the construction of the ensemble models and the determination of class-specific decision thresholds. Even though the test set was maintained strictly independent from these processes, this approach may increase the risk of overfitting to the validation set.

Additionally, all experiments were conducted using a single seed value. This condition limits the evaluation of performance variability that might emerge across different training iterations. The models were evaluated exclusively on the PTB-XL dataset. Consequently, the generalizability of the obtained results to noisier ECG records acquired from diverse centers and encountered in real-world clinical environments necessitates validation on independent datasets.

\section{CONCLUSION}

This study compares convolutional, recurrent, state-space, time-frequency-based transformer, and pre-trained ECG-specific foundation model approaches for multi-label 12-lead ECG classification on the PTB-XL dataset within a common experimental framework. On the independent test set, ECGFounder demonstrated the strongest overall performance among the individual models, while the heterogeneous ensemble methods achieved superior results compared to the individual models across the primary performance metrics. The Stacking Ensemble approach attained the highest Macro AUROC, Macro AUPRC, and Macro F1 scores, whereas validation-weighted soft voting yielded the highest Subset Accuracy. In contrast, CWT-ViT-KAN underperformed compared to the other individual architectures under the current experimental conditions. These results indicate that heterogeneous ensembles, which leverage the complementary characteristics of diverse representation learning approaches, can offer potential advantages in multi-label ECG classification. Although the final performance metrics were obtained on the pre-split PTB-XL test set, external validation on independent datasets and noisy clinical ECG records under real-world conditions is crucial to demonstrate the broader clinical generalizability of these findings.

\end{document}